\documentstyle[aps,prb,preprint]{revtex}
\tighten
\newcommand{\beq}{\begin{eqnarray}}
\newcommand{\eeq}{\end{eqnarray}}
\begin{document}
\draft
\input epsf.sty

\title
{Landau theory of bi-criticality in a random quantum rotor system}
\author{Denis Dalidovich and Philip Phillips}
\vspace{.05in}

%
\address
{Loomis Laboratory of Physics\\
University of Illinois at Urbana-Champaign\\
1100 W.Green St., Urbana, IL, 61801-3080}

%
\maketitle
\begin{abstract}

We consider here a generalization of the random quantum rotor
model in which each rotor is characterized by an M-component
vector spin. We focus entirely on the case not considered
previously, namely when the distribution of exchange interactions
has non-zero mean.  Inclusion of non-zero mean permits
ferromagnetic and superconducting phases for $M=1$ and $M=2$,
respectively.  We find that quite generally, the Landau theory for
this system can be recast as a
 zero-mean problem in the presence of a magnetic field.
Naturally then, we find that a Gabay-Toulouse line exists for
$M>1$ when the distribution of exchange interactions has non-zero
mean. The solution to the saddle point equations is presented in
the vicinity of the bi-critical point characterized by the
intersection of the ferromagnetic (M=1) or superconducting (M=2)
phase with the paramagnetic and spin glass phases. All transitions
are observed to be second order.  At zero temperature, we find
that the ferromagnetic order parameter is non-analytic in the
parameter that controls the paramagnet/ferromagnet transition in
the absence of disorder.  Also for M=1, we find that replica
symmetry breaking is present but vanishes at low temperatures.  In
addition, at finite temperature, we find that the qualitative
features of the phase diagram, for $M=1$, are {\it identical} to
what is observed experimentally in the random magnetic alloy
$LiHo_xY_{1-x}F_4$.
\end{abstract}


\columnseprule 0pt
\narrowtext

{  }

\section{Introduction}

Transport in granular metals is mediated by activated transport
among the metallic grains.  In granular superconductors composed
of spatially separated metallic grains, such single particle
charging events ultimately lead to a destruction of phase locking
between the grains. This state of affairs obtains because the
particle number, $n$, and phase, $\theta$, associated with each
grain are conjugate variables.  If two grains phase lock, the
resultant infinite uncertainty in the particle number leads
necessarily to single particle charging.  Should the single
particle charging energy sufficiently exceed the Josephson
coupling energy between grains, superconductivity is
quenched\cite{anderson}.

The simplicity of the physics underlying the quantum phase
transition in an array of superconducting islands implies that
the resultant Hamiltonian, \beq\label{HJ}
H=-E_C\sum_i\left(\frac{\partial}{\partial\theta_i}\right)^2-
\sum_{\langle i,j\rangle} J_{i,j}\cos(\theta_i-\theta_j), \eeq is
characterized by only two parameters: 1) a charging energy,
$E_C$, and 2) a Josephson coupling energy, $J_{i,j}$.  In
writing Eq. (\ref{HJ}), we assumed that the islands occupy regular
sites on a 2D lattice and only nearest neighbor, $\langle
i,j\rangle$, Josephson coupling is relevant. For the ordered case
in which $J_{\langle i,j\rangle}=J_0$, the superconductor-insulator
transition is well-studied\cite{doniach} as the parameter
$E_C/J_0$ increases.  In sufficiently disordered systems,
however, the Josephson energies are not all equal and in fact can
be taken to be random.

We are concerned in this paper with the case in which the
Josephson energies are random and characterized by a Gaussian
distribution \beq P(J_{ij})=\frac{1}{\sqrt{2\pi
J^2}}\exp{\left[-\frac{(J_{ij}-J_0)^2}{2J^2}\right]} \eeq with
non-zero mean, $J_0$.  While random Josephson systems have been
treated previously\cite{sachdev,huse,bm}, such studies have
focused predominantly on the zero mean case in which $J_0=0$. This
limit differs fundamentally from the non-zero mean case, because
the superconducting phase exists only when $J_0\ne 0$. Hence, for
$J_0=0$, there is an absence of an ordering transition.
Nonetheless, the zero-mean case is still of physical interest
because as Read, Sachdev, and Ye (RSY)\cite{sachdev} as well as
Miller and Huse\cite{huse} have shown, a zero temperature
transition from a quantum spin-glass to a paramagnet occurs as the
strength of the quantum fluctuations increases.  To put the random
Josephson model in the context of the work of RSY, it is expedient
to introduce the change of variables \beq {\bf S}_i=(\cos\theta_i,
\sin\theta_i). \eeq
 The
resultant Josephson Hamiltonian
 \beq\label{hspin}
H=-E_C\sum_i\left(\frac{\partial}{\partial\theta_i}\right)^2+
\sum_{\langle i,j\rangle}J_{i,j}{\bf S}_i\cdot{\bf S}_j
 \eeq
is recast as a 2-component (M=2) interacting spin problem with
random magnetic interactions.  This model is easily generalizable
to describe interactions among any M-component spin operator (or
quantum rotor) in the group $O(M)$. M=1 corresponds to Ising
spins and is relevant to random magnetic\cite{rosenbaum} systems such as
LiHo$_{0.167}$Y$_{0.833}$F$_4$ whereas the M=3 limit is applicable
to spin fluctuations in the cuprates\cite{haldane}.  In this paper
we focus primarily on the M=1 and M=2 cases in which the ordered
phases correspond to a ferromagnet and a superconductor,
respectively.

Two types of fluctuations control the transitions between the
phases: 1) dynamic quantum fluctuations arising from the charging
energy and 2) static fluctuations induced by the disorder. For
non-zero mean, three phases (spin-glass, paramagnet, ferromagnet
(M=1) or superconductor (M=2)) are expected to meet at a
bi-critical point.  Experimentally, a bi-critical point is
anticipated whenever $J_0$, $J$ and $E_C$ are on the same order of
magnitude. It is the physics at this bi-critical point that we
focus on in this paper. To discriminate between these phases, we
distinguish between thermal averages, $\langle ...\rangle$ and
averages over disorder, $[...]$.  In the superconductor (M=2) or
ferromagnetic phases (M=1), the disorder and thermal average of
the local spin operator, $[\langle S_{i\nu}\rangle]\ne 0$, is
non-zero.  In the spin glass phase, the thermal average $\langle
S_{i\nu}\rangle\ne 0$, while $[\langle S_{i\nu}\rangle]=0$.  At
zero temperature, quantum fluctuations and static disorder
conspire to lead to a vanishing of the static moment in the
paramagnetic phase, that is, $\langle S_{i\nu}\rangle=0$.  RSY
have performed an extensive study of the spin-glass/paramagnetic
boundary using the replica formalism\cite{sk} in the case of zero
mean in which the purely ordered phase is absent.  We adopt this
formalism here in our analysis of the bi-critical region. The only
prior study on the non-zero mean case is that of Hartman and
Weichman\cite{hw} who studied numerically the spherical limit,
$M\rightarrow\infty$, of the quantum rotor Hamiltonian.  In their
study, they found that the spin-glass phase is absent in the
$M\rightarrow\infty$ limit for $d=2$.   In the present work, we
will not address the issue of dimensionality because we limit
ourselves to a mean-field description in which all fields are
homogeneous in space.  Our results then are valid above some upper
critical dimension that can only be determined by including
fluctuations around the mean-field solution.

This paper is organized as follows.  In section II, we use the
replica formalism to average over the disorder explicitly and
obtain the effective Landau free-energy functional.  We show that
the leading terms in the Landau action resulting from the non-zero
mean are analogous to those arising from an external magnetic
field in the zero-mean problem.  As a result, the appropriate
saddle-point equations can be solved using a direct analogy to the
zero-mean problem in the presence of a magnetic  field.  Explicit
criteria are presented in Sec. III for the stability of each of
the phases in the vicinity of bi-criticality.  We construct the
phase diagram at finite temperature and find excellent agreement
with the experimental results of Reich\cite{rosenbaum}, et. al. on
the magnetic system $LiHo_xY_{1-x}F_4$.  The explicit solution for
$M\ge 2$ is presented in the last section with a special emphasis
on the Gabay-Toulouse\cite{gt} line. In section IV we analyze the
possibility of replica symmetry breaking along the de
Almeida-Thouless\cite{at} line in the M=1 case.

\section{Landau Action}

Central to the construction of a Landau theory of the bi-critical region
is the free-energy functional.  For a quantum mechanical system, this
is obtained by explicitly including in the partition function,
\beq\label{par}
Z=Tr\left(e^{-\beta H}\right)=Tr\left[e^{-\beta H_0}\widehat{T}
\exp\left[-\int_0^\beta H_1(\tau)d\tau\right]\right]
\eeq
the time evolution according to a reference system. For this problem,
the Hamiltonian for free quantum rotors
\beq
H_0= -E_C\sum_i\left(\frac{\partial}{\partial\theta_i}\right)^2
\eeq
describes our reference system.  The perturbation
\beq
H_1(\tau)=e^{H_0\tau}\sum_{\langle i,j\rangle}J_{i,j}{\bf S}_i\cdot{\bf S}_je^{-H_0\tau}
 \eeq
corresponds to the random magnetic interactions.  The trace in
the partition function is evaluated over the complete set of quantum
rotor states.  A primary hurdle in evaluating the
partition function is the average over the random spin interactions.
 It is now
standard to perform this average\cite{bm} by replicating the spin
system $n$ times and using the identity,
\beq
\ln [Z]=\lim_{n\rightarrow 0}\frac{[ Z^n]-1}{n} \eeq to
obtain the partition function, $Z$.  We first must then evaluate
$\langle Z^n\rangle$. Formally, the replicated partition function
is defined through Eq. (\ref{par}) by replacing $H_0$ and $H_1$ by
their replicated equivalents \beq H_i^{\rm eff}=\sum_{a=1}^n H_i^a
\eeq where the superscript indexes the individual replicas and
$i=0,1$.  Within the replica formalism, the average over the
random interactions with the Gaussian distribution gives rise to a
quartic spin interaction.  This quartic spin interaction is easily
reduced to a quadratic interaction by use of the
Hubbard-Stratanovich transformation.  Let us define the Fourier
transforms \beq {\bf S}^a(k,\tau)=\frac{1}{\sqrt{N}}\sum_i {\bf
S}_i^a(\tau) e^{i{\bf k\cdot R_i}} \eeq of the local spin operator
for each site and the corresponding transform of the
nearest-neighbour interaction \beq J(k)=J\sum_{\langle
ij\rangle}e^{\bf k\cdot r_{ij}}=2J\sum_{i=1}^d cosk_ia \eeq The
replicated partition function now takes on the form,
\beq
[ Z^n]=Z_0^n\int D\Psi D Qe^{-F_{\rm eff}}[Q,\Psi]
\eeq
 where the effective Free-energy
\beq\label{free}
F_{\rm eff}[\Psi, Q]&=&\sum_{a,k}\int_0^\beta d\tau\Psi^a_{\mu}(k,\tau)
\cdot(\Psi^a_{\mu}(k,\tau))^\ast
+\sum_{a,b,k,k'}
Q_{\mu\nu}^{ab}(k,k',\tau,\tau')[Q_{\mu\nu}^{ab}(k,k',\tau,\tau')]^\ast\nonumber\\
&&-\ln\left[\langle \widehat{T}\exp\left(2\int_0^\beta d\tau\sum_{a,k}\sqrt{J_0(k)}\Psi^a_\mu(k,\tau)
S^a_\mu(-k,\tau)\right.\right.\nonumber\\
&&\left.\left.+\sum_{a,b,k,k'}\int_0^\beta\int_0^\beta d\tau d\tau'\sqrt{2J(k)J(k')}
Q_{\mu\nu}(k,k',\tau,\tau')S_\mu^a(k,\tau)S_\nu^b(-k',\tau')\right)\rangle_0
\right]
\eeq
is now a functional of the auxilliary fields $Q$ and $\Psi$ which
appear upon use of the Hubbard-Stratanovich transformation to decouple
the quartic spin term proportional to $J^2(k)$ and the quadratic
term scaling as $J_0(k)$, respectively.  In writing this expression,
we used the Einstein convention where repeated spin (but not replica) indices are summed over,
$Z_0=Tr[\exp(-\beta H_0)]$, and
\beq
\langle A\rangle_0=\frac{1}{Z_0^n}Tr\left(e^{-\beta H_0^{\rm eff}} A\right)
\eeq

The fields $Q$ and $\Psi$ play fundamentally different roles. The
proportionality of the $\Psi$ field to the mean of the
distribution implies that this field determines the ordering
transition (superconducting for M=2 or ferromagnetic for M=1).
This can be seen immediately upon differentiating the free energy
with respect to $\Psi$.  The self-consistent condition is that
\beq \Psi^a_\mu(k,\tau)=\langle S^a_\mu(k,\tau)\rangle \eeq Hence,
a non-zero value of $\Psi^a_\mu(k,\tau)$ implies ordering.  It is
for this reason that $\Psi$ functions as the order parameter for
the ferromagnetic or superconducting phase within Landau theory.
Likewise, differentiation of the free energy with respect to $Q$
reveals that \beq Q_{\mu\nu}^{ab}(k,k',\tau,\tau')=\langle
S_\mu^a(k,\tau)S_\nu^b(k',\tau')\rangle \eeq is the
self-consistency condition for the $Q$ matrices. For quantum spin
glasses, it is the diagonal elements of the Q-matrix \beq
D(\tau-\tau')=\lim_{n\rightarrow 0}\frac{1}{Mn}\langle
Q^{aa}_{\mu\mu}(k,k',\tau,\tau')\rangle \eeq in the limit that
$|\tau-\tau'|\rightarrow\infty$ that serves as the effective
Edwards-Anderson spin-glass order parameter\cite{sachdev,bm}
within the Landau theory.  A disclaimer is appropriate here as $D$
is non-zero in the spin-glass as well as the paramagnet phases.
However, its behaviour is sufficiently different in the three
phases: in the paramagnetic and ferromagnetic (superconducting)
phases, as we will show, $D(\omega)$ still has the form
$\sqrt{\omega^2+\Delta^2}$. The gap, $\Delta$, vanishes at the
transition to the spin-glass phase giving rise to the long-time
behavior of $D(\tau)$.

Before we analyze the form of the gap, we must obtain the
effective Landau action. The goal here is to obtain a polynomial
functional of $Q$ and $\Psi$ from which a saddle point analysis
can be performed. We proceed in the standard way and perform a
cumulant expansion on the free energy.  For the case of zero-mean,
this procedure is documented quite closely in RSY.  In addition to
the terms containing the $Q$ matrices studied by RSY, the non-zero
mean case will contain powers of $\Psi^a_\mu$ as well as cross
terms.  The resultant action must contain, of course, only even
powers of $\Psi_\mu^a$. For an analysis of the bi-critical point,
it is sufficient to retain quadratic and quartic terms in
$\Psi_\mu^a$.  Terms of this kind are completely analogous to
those derived previously by Doniach\cite{doniach}.  Of the cross
terms, the simplest are of the form
$\Psi^a_\mu\Psi_\nu^bQ_{\mu\nu}^{ab}$,
$\Psi^a_\mu\Psi_\nu^aQ_{\mu\nu}^{aa}$, and
$\Psi_\mu^a\Psi_\mu^aQ_{\nu\nu}^{aa}$. We have confirmed
explicitly that retension of the latter two terms in which only
one replica index occurs leads only to the renormalization of
various coupling constants and minor modification of the phase
boundaries near the bi-critical region.  Hence, we do not consider
such terms. Likewise, we do not retain terms of the form
$Q^{aa}Q^{bb}$ because this term vanishes in the limit
$n\rightarrow 0$.

We find then that the effective action \beq\label{action}
\cal{A}&=&\frac{1}{t}\int d^dx\left\{\frac{1}{\kappa}\int
d\tau\sum_a
\left(r+\frac{\partial}{\partial\tau_1}\frac{\partial}{\partial\tau_2}
\right)Q_{\mu\mu}^{aa}(x,\tau_1,\tau_2)|_{\tau_1=\tau_2=\tau}\right.\nonumber\\
&&+\frac12\int d\tau_1 d\tau_2\sum_{a,b}\left[\nabla
Q^{ab}_{\mu\nu}(x,\tau_1,\tau_2)\right]^2\nonumber\\
&&-\frac{\kappa}{3}\int d\tau_1 d\tau_2
d\tau_3\sum_{a,b,c}Q^{ab}_{\mu\nu}(x,\tau_1,\tau_2)
Q^{bc}_{\nu\rho}(x,\tau_2,\tau_3)Q^{ca}_{\rho\mu}(x,\tau_3,\tau_1)\nonumber\\
&&\left.+\frac12\int
d\tau\sum_a\left[uQ^{aa}_{\mu\nu}(x,\tau,\tau)Q^{aa}_{\mu\nu}(x,\tau,\tau)
+vQ^{aa}_{\mu\mu}(x,\tau,\tau)Q^{aa}_{\nu\nu}(x,\tau,\tau)\right]\right\}\nonumber\\
&&+\frac{1}{2g}\int d^dx\left\{ d\tau \sum_a\Psi_\mu^a(x,\tau)
\left[-\gamma-\frac{\nabla^2}{2}\right]\Psi_\mu^a(x,\tau)\right.\nonumber\\
&&\left.+\int
d\tau\sum_a\frac{\partial}{\partial\tau_1}\Psi_\mu^a(x,\tau_1)
\frac{\partial}{\partial\tau_2}\Psi_\mu^a(x,\tau_2)|_{\tau_1=\tau_2=\tau}
+\frac{\zeta}{2}\int
d\tau\sum_a\left[\Psi_\mu^a(x,\tau)\Psi_\mu^a(x,\tau)\right]^2\right\}\nonumber\\
&&-\frac{1}{\kappa t g}\int d^d x\int d\tau_1
d\tau_2\sum_{a,b}\Psi_\mu^a(x,\tau_1)
\Psi^b_\nu(x,\tau_2)Q_{\mu\nu}^{ab}(x,\tau_1,\tau_2)\+\cdots \eeq
contains three types of terms. 1) The terms that depend purely on
the Q-matrices, as in the terms in the first curly bracket in Eq.
(\ref{action}), have been studied previously by RSY in the context
of quantum spin-glass/paramagnet transition. 2) The
$\Psi$-dependent terms in the second curly bracket give rise to
the ferromagnet or superconducting phases.  They are controlled by
the parameter $\gamma$.  3) The competition between these two
transitions is mediated by the last term in Eq. (\ref{action}). To
re-iterate, other cross-terms exist--for example the term noted
previously.  However, such terms have no bearing on the critical
region.  It suffices then to truncate the action at the level of
Eq. (\ref{action}). The last term in Eq. (\ref{action}) bears a
strong resemblance to the term that appears in the Landau action
for the zero-mean problem in the presence of a magnetic  field,
$h$.  In fact, we can obtain the corresponding term from Eq.
(\ref{action}) by the transformation $\Psi^a_\mu\rightarrow
h\delta_{\mu,1}\sqrt{\kappa g/2t}$. This mapping is not
unexpected. In the non-zero mean problem, ordered spins create an
effective ``magnetic field'' that acts as a source term for the
$Q$-matrices. This analogy is particularly powerful and serves as
a useful check as to the validity of our saddle-point equations.
Further, this mapping is true even in the classical case.

A word on the coupling constants is in order. The parametrization
of the action in terms of the coupling constants $\kappa$, $t$,
and $g$ was obtained by appropriately rescaling the fields $Q$ and
$\Psi$ as well as the space and time coordinates. Fundamentally,
$g$ is a function of $E_C$ and $J_0$, while $\kappa$ and $t$ are
functions of $E_C$ and $J$ only.  $u$, $v$, and $\zeta$ are
related to a four-point spin correlation function evaluated in the
long-time limit as discussed by RSY.  In the zero-mean case, the
phase diagram demarcating spin-glass and paramagnetic stability is
determined by the parameter $r$ which determines the strength of
the quantum fluctuations.  When these fluctuations exceed a
critical value, that is $r>r_c$, a transition to a paramagnetic
phase occurs\cite{sachdev}. In the problem at hand, two
parameters, $r$ and $\gamma$, determine the phase diagram.  The
coupling constant, $\gamma$, is directly related to the parameter
$E_C/J_0$ and $E_C/J$ as well. The latter dependence arises as a
result of the removal of the quadratic term, \beq \int d^dx
d\tau_1d\tau_2\sum_{a,b}[Q_{\mu\nu}^{ab}(x,\tau_1,\tau_2)]^2,\nonumber\\
\eeq by the transformation\cite{sachdev}
 $Q\rightarrow Q-C\delta^{ab}\delta_{\mu}(\tau_1-\tau_2)$.
From microscopic considerations, it follows that $\kappa^2 t/2=1$.
This is an important simplification because we will show that
the ferromagnet/paramagnet
boundary is determined by the line $\gamma=2\Delta/\kappa^2 t=\Delta$.

\section{Saddle Point Analysis: Phase Diagram}

In terms of the frequency-dependent order parameters, \beq
\Psi_\mu(k,\omega)=\frac{1}{\beta}\int_0^\beta d\tau
\Psi_\mu^a(k,\tau) e^{-i\omega\tau} \eeq and \beq
Q_{\mu\nu}^{ab}(k_1,k_2,\omega_1,\omega_2)&=&\frac{1}{\beta^2}\int_0^\beta\int_0^\beta
d\tau_1 d\tau_2 Q_{\mu\nu}^{ab}(k_1,k_2,\tau_1,\tau_2)
e^{-i\omega_1\tau_1-i\omega_2\tau_2}, \eeq the three phases in our
problem form under the following conditions. For the paramagnetic
phase near the bi-critical point, the saddle-point equations are
satisfied when $\Psi^a_\mu=0$ but as RSY have shown \beq
Q_{\mu\nu}^{ab}(k,\omega_1,\omega_2)=(2\pi)^d\beta\delta_{\mu\nu}\delta^{ab}\delta^d(k)
\delta_{\omega_1+\omega_2,0}D(\omega_1). \eeq We have retained the
ansatz for the $Q$-matrix used by RSY for the paramagnetic phase.
Similarly, the ordering parameter also vanishes in the spin-glass
phase, $\Psi_\mu=0$ and for the replica-symmetric solution, \beq
Q_{\mu\nu}^{ab}(k,\omega_1,\omega_2)=(2\pi)^d\delta^d(k)\delta_{\mu\nu}\left[\beta
D(\omega_1)
\delta_{\omega_1+\omega_2,0}\delta^{ab}+\beta^2\delta_{\omega_1,0}\delta_{\omega_2,0}q^{ab}
\right]. \eeq The possibility of a replica broken-symmetry
solution will also be explored by including terms of $O(Q^4)$ in
the Landau action. In the ordered phase (that is ferromagnetic or
superconductor) both $\Psi$ and $Q$ are nonzero. As ferromagnetism
has generally been studied with a frequency-independent order
parameter, we explore a static ansatz of the form \beq\label{psi}
\Psi_\mu^a(k,\omega)=(2\pi)^d\delta^d(k)\beta
\delta_{\omega,0}\delta_{\mu,1}\psi \eeq
 for $\Psi^a_\mu$
and
\beq\label{qm}
Q_{\mu\nu}^{ab}(k,\omega_1,\omega_2)=(2\pi)^d\delta^d(k)\delta_{\mu\nu}\left[\beta\bar{D}(\omega_1)
\delta_{\omega_1+\omega_2,0}\delta^{ab}+\beta^2\tilde{q}\delta_{\omega_1,0}
\delta_{\omega_2,0}\delta^{ab}
+\beta^2\delta_{\omega_1,0}\delta_{\omega_2,0}q^{ab} \right] \eeq
for the Q-matrices. Our explicit inclusion of $\tilde{q}$ as the
$\omega=0$ diagonal element of the Q-matrices implies that we can
redefine $D(\omega)=\bar{D}(\omega)+\beta
\tilde{q}\delta_{\omega,0}$.  Consequently, we can set
$\bar{D}(\omega)=0$ and assume that $q^{ab}$ is purely
off-diagonal.  The replica-symmetric solution corresponds to
$q^{ab}=q$ for all $a\ne b$.  Initially, we will explore only
this case.

\subsection{M=1: Ferromagnetic order}

We now specialize to the $M=1$ ferromagnetic case as crucial
differences can occur with the analysis for $M>1$.  If we
substitute Eqs. (\ref{psi}) and (\ref{qm}) into the Landau action,
we obtain a free-energy density of the form \beq\label{fed}
\frac{\cal F}{n}&=&\frac{1}{\beta\kappa t}\sum_{\omega\ne
0}(\omega^2+r)\bar{D}(\omega)\ +\frac{r\tilde{q}}{\kappa
t}-\frac{\kappa}{3\beta t}\sum_{\omega\ne
0}\bar{D}^3(\omega)\nonumber\\
&&+\frac{u+v}{2t}\left(\tilde{q}+\frac{1}{\beta}\sum_{\omega\ne
0}\bar{D}^2(\omega)\right)^2
+\frac{1}{2g}\left(-\gamma\psi^2+\frac{\zeta
\psi^4}{2}\right)\nonumber\\
&&-\frac{\kappa\beta^2}{3t}\left(\tilde{q}^3+3\tilde{q}\frac{Tr
q^2}{n}+\frac{Tr q^3}{n}\right) -\frac{\beta\psi^2}{\kappa g
t}\left(\tilde{q}+\frac{\sum_{ab} q^{ab}}{n}\right). \eeq
Formally, the free-energy density as defined here is the
disorder-averaged free energy per replica per spin component.  The
last term in Eq. (\ref{fed}) arises from the cross term and
generates off-diagonal components of the Q-matrix. For $\psi=0$,
this term is absent and we generate the solution of RSY. We obtain
the saddle point equations by differentiating Eq. (\ref{fed}) with
respect to $q$, $\tilde q$, $\bar D$, and $\psi$ in the
$n\rightarrow 0$ limit. All of these quantities can be simplified
once the appropriate gap parameter, \beq
\Delta^2=r+\kappa(u+v)\left(\tilde{q}+
\frac{1}{\beta}\sum_{\omega\ne 0}\bar{D}(\omega)\right) \eeq is
identified.  The difference with the corresponding quantity in the
work of RSY is the presence of the $\tilde{q}$ term in the gap
which includes the contribution arising from the cross term in the
Landau action.  From the derivative equation with respect to $\bar
D(\omega)$, we find that \beq
\bar{D}(\omega)=-\frac{1}{\kappa}\sqrt{\omega^2+\Delta^2}. \eeq
The constraint, $\Delta\ge 0$ is crucial for the stability of the
free energy density and the minus sign in the above expression
ensures that $D(\tau)>0$.  The saddle-point equations for $q$ and
$\tilde{q}$ yield that \beq q=\frac{\psi^2}{2\kappa g\Delta} \eeq
and \beq \tilde{q}=\frac{\psi^2}{2\kappa
g\Delta}-\frac{\Delta}{\kappa \beta}. \eeq As expected, these
equations form the basis for the bulk of our analysis and are
identical to the RSY saddle-point equations with $h$ replaced by
$\psi$.  Despite this mapping, $h$ and $\psi$ do serve different
roles in the zero and non-zero mean theories.  In the former, $h$
is an external adjustable scalar quantity without any critical
properties, whereas $\psi$ plays the role of an order parameter in
the non-zero mean case and must be treated on equal footing with
$Q$.

The saddle-point equation for $\psi$ \beq
\psi\left[-\gamma+\Delta+\zeta \psi^2\right]=0 \eeq implies that
aside from the trivial solution $\psi=0$, the non-trivial solution
corresponds to \beq\label{order}
\psi^2=\frac{1}{\zeta}\left(\gamma-\Delta\right). \eeq In
obtaining this equation, we used the fact that $\kappa^2 t/2=1$.
If we specialize to low temperatures, that is low relative to the
gap ($T<\Delta$), the sum in the gap equation can be evaluated
\beq\label{sum}
\frac{1}{\beta}\sum_\omega\sqrt{\omega^2+\Delta^2}=\frac{\Lambda_\omega^2}{2\pi}+\frac{\Delta^2}{4\pi}\ln\frac{\Lambda_\omega^2}{\Delta^2}
\eeq and the resultant self-consistent equation takes the form
\beq\label{gap}
\Delta^2=r-r_c-\frac{u+v}{4\pi}\Delta^2\ln\frac{\Lambda_\omega^2}{\Delta^2}
+\frac{u+v}{2g\zeta}\left(\frac{\gamma}{\Delta}-1\right). \eeq We
have introduced \beq\label{rc}
r_c=(u+v)\frac{\Lambda_\omega^2}{2\pi} \eeq and the cut-off
$\Lambda_\omega$ is determined by the energy scale at which
zero-point quantum fluctuations become important.  The natural
cut-off for such fluctuations in this problem is $E_C$, the
charging energy.

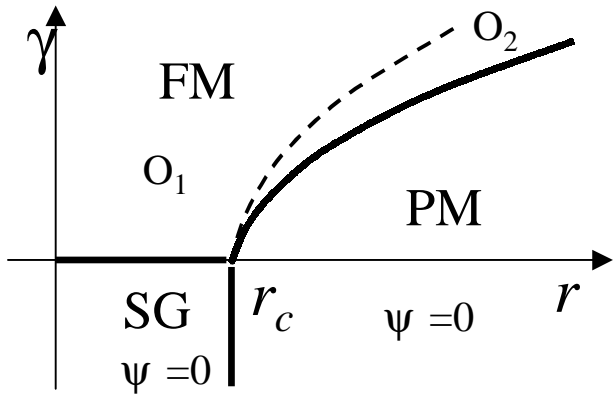
\begin{figure}
\caption{Zero-temperature phase diagram demarcating the regions of
ferromagnetism (FM), paramagnetism and spin-glass behaviour for
M=1.  The parameters $\gamma$ and $r$ are determined by the
dynamical quantum fluctuations and the static disorder. The curve
separating the ferromagnet from the paramagnet scales roughly as
$\gamma\approx \sqrt{r-r_c}$ up to logarithmic factors. Regions
$O2$ and $O1$ are distinguished by the magnitude of the spin-wave
gap, $\Delta$, as well as the magnitude of the order parameter
$\psi$, which is non-zero only in the ferromagnetic phase. The
transition between these two regions occurs when
$\gamma-\gamma_c\approx (r-r_c)^{3/2}$, where $\gamma_c$ and $r_c$
are determined by Eqs. (\protect\ref{gamline}) and
(\protect\ref{rc}), respectively.} \label{fig1}
\end{figure}

The essential physics of the bi-critical point is contained in
Eqs. (\ref{order}) and (\ref{gap}).  From Eq. (\ref{order}), it is
clear that the ferromagnetic phase exists only when
$\gamma\ge\Delta$. This ensures that $\psi^2>0$ in the
ferromagnetic phase. The vanishing of the order parameter $\psi$
signifies a termination of the ferromagnetic phase. For $\gamma\ne
0$ and $\Delta\ne 0$, the only line along which $\psi^2=0$
corresponds to $\gamma=\Delta$. If we substitute this condition
into the gap equation, we find that the critical line separating
the ferromagnet from the paramagnet at low temperatures
($T\ll\Delta$) is given by \beq\label{gamline}
\gamma=\Delta=\left(\frac{4\pi(r-r_c)}{(u+v)
\ln\frac{\Lambda_\omega^2}{r-r_c}}\right)^{\frac{1}{2}} \eeq and
is depicted in Fig. (\ref{fig1}). In obtaining this equation, we
assumed that $|r-r_c|\ll 1$ and $\gamma\ll 1$.  The bi-critical
point corresponds to $r=r_c$ and $\gamma=0$.  At this point, the
gap vanishes as does $\psi$. The essential non-trivial nature of
this result is the non-analytic dependence of $\gamma$ on the
quantum fluctuation parameter, $r-r_c$.

In the vicinity of the bi-critical point, distinct regimes
partition the ferromagnetic phase that are determined by the
magnitude of $\Delta$ and $\psi$. This behavior is shown in Fig.
(\ref{fig1}).  The regions $O1$ and $O2$ are distinguished by
their distance from the line $\gamma=\Delta$.  The transition
between $O1$ and $O2$ occurs when \beq\label{gaml2}
\gamma-\left(\frac{4\pi(r-r_c)}{(u+v)
\ln\frac{\Lambda_\omega^2}{r-r_c}}\right)^{\frac{1}{2}}\approx
\frac{(r-r_c)^{\frac32}}{\ln\frac{\Lambda_\omega^2}{r-r_c}} \eeq
Because $r-r_c$ is much less than unity, algebraic dependence in
Eq. (\ref{gtt2}) of the form, $(r-r_c)^{3/2}$, implies that the
$O2$ region is quite narrow. Within this narrow region, the gap is
well approximated by Eq. (\ref{gamline}) and the value of $\psi$
is given by \beq
\psi^2=\frac{1}{\zeta}\left(\gamma-\left(\frac{4\pi(r-r_c)}{(u+v)
\ln\frac{\Lambda_\omega^2}{r-r_c}}\right)^{\frac{1}{2}}\right)\quad\quad
{\rm region}\quad O2 \eeq In the bulk of the ferromagnetic phase,
region $O1$, the $\gamma$-dependent terms must be retained in Eq.
(\ref{gap}) to accurately describe the gap,
$\Delta=\gamma-\zeta\psi^2$, with the ordering parameter given by
\beq\label{psi2} \psi^2=\frac{2g}{u+v}\left[\frac{u+v}{2\pi}
\gamma^3\ln\frac{\Lambda_\omega}{\gamma}-\gamma(r-r_c)\right]\quad\quad
{\rm region}\quad O1 \eeq Hence, at the point $r=r_c$, we find a
strong non-analytic dependence,
$\psi\propto\gamma^{3/2}\sqrt{\ln\Lambda_\omega/\gamma}$, on the
coupling constant $\gamma$.  This is particularly important
because it signifies that even at the mean-field level, deviations
from the standard square-root dependence are present in the
current theory.  This result is fundamentally tied to the
logarithmic dependence induced by the frequency summations and is
caused by the zero-point quantum fluctuations.  However, if we
extrapolate our results to the regime, $r-r_c\approx O(1)$, the
second term in Eq.(\ref{psi2}) dominates and we do recover that
$\psi\propto\gamma^{1/2}$ in agreement with the expectation from
standard mean-field Landau theories. Paramagnetic and spin-glass
behaviour obtain whenever $\gamma<\Delta$.  In this regime, the
non-trivial solution for $\psi$ no longer holds and $\psi=0$ is
the only valid solution.  In this limit, our solution for $\Delta$
is identical to that of RSY and all of their results are
recovered.  For example, consider the free energy density (in
units of $\kappa^2 t/2=1$),
 \beq\label{F0}
\frac{\cal{F}}{n}&=&\frac{\Lambda_\omega^4}{2\pi}\left(\frac16+\frac{u+v}{8\pi}\right)
-{\Lambda_\omega^2(r-r_c)\over 4\pi}-{(r-r_c)^2\over
4(u+v)}+{\gamma^2(r-r_c)\over 2(u+v)}
-\frac{\gamma^4}{8\pi}\ln{\Lambda_\omega\over\gamma}+\cdots\nonumber\\
&=&{\cal
F}_0-\frac{\gamma^4}{8\pi}\ln{\frac{\Lambda_\omega}{\gamma}}+
{\gamma^2(r-r_c)\over 2(u+v)}+\cdots
 \eeq
 in region $O1$.  This quantity is obtainable from Eq.
(\ref{fed}) once the saddle point solutions for region $O1$ are
used and only the leading terms in $\gamma$ and the cross-term
are retained.  When $\gamma=0$, this expression is identical to that of
RSY in the spin-glass phase. As anticipated, the leading $\gamma$ dependence in the
free-energy density is non-analytic.  This behaviour originates
from the non-analytic behaviour of the order parameter $\psi$ on
$\gamma$. We will see that this non-analyticity does not survive
for $M>1$ above and below the Gabay-Toulouse line.

Extending these results to finite temperatures, $T\gg\Delta$,
simply requires the evaluation of the frequency summation over
frequencies in the gap equation for $T\gg\Delta$. Using the result
in Eq. (2.l3) of RSY, we obtain a self-consistent condition for
the gap \beq\label{eq40}
\Delta^2=r-r_c(T)-(u+v)T\Delta-\frac{u+v}{2\pi}\Delta^2\ln
\frac{\Lambda_{\omega}}{T}+\frac{u+v}{2g\zeta\Delta}(\gamma-\Delta)
\eeq in the regime $T\gg\Delta$ where we have defined
$r_c(T)=r_c-(u+v)\pi T^2/3$.  Recall that the transition between
the ferromagnetic and paramagnetic phases occurs when
$\gamma=\Delta$. Hence, in the units chosen here, this condition
simplifies to $\gamma=\Delta(T)$.  For $T\ll\sqrt{r-r_c(T)}$, the
boundary for the paramagnetic/ferromagnetic state remains
unchanged from the $T=0$ results discussed above.  However, for
$T\gg\Delta$, two distinct regimes
\beq \gamma=\Delta(T)=\left\{
\begin{array}{lll} \frac{r-r_c(T)}{\sqrt{(u+v)}T} & \quad
\sqrt{r-r_c(T)}\ll T & \quad {\rm region \quad O3}\\
\sqrt{\frac{2\pi^2T^2}{3\ln\frac{\Lambda_\omega}{T}}} & \quad
\sqrt{r-r_c}\ll T & \quad {\rm region \quad O4}
\end{array}\right.
\eeq
 emerge depending on the magnitude of the thermal
fluctuations. These regimes are depicted in Fig.(\ref{fig2}a). The
crossover between these two regions occurs when
$\gamma\propto\sqrt{r_c-r}$. In Fig. (\ref{fig2}a), $r_c-r>0$. The
temperature \beq T_0=\sqrt{\frac{3(r_c-r)}{\pi(u+v)}} \eeq is
denoted explicitly in Fig. (\ref{fig2}a) as this is the lowest
temperature at which region $O3$ obtains. Immediately below region
$O3$ where $\gamma-\Delta_0(T)\propto(r-r_c(T))^2/T$ and to the
right of $O4$ where $\gamma-\Delta_0(T)\propto
T^3/\sqrt{\ln\Lambda_\omega/T}$, a transition to a new region
occurs in which
 the gap takes the
form, \beq\label{ftpsi}
\Delta=\gamma-\frac{2g\zeta}{u+v}\left[{(u+v)\gamma^3\over
2\pi}\ln\frac{\Lambda_\omega}{T}
+(u+v)\gamma^2T-(r-r_c(T))\gamma\right]=\gamma-\zeta\psi^2(T) \eeq
In this region, denoted as $O5$ in Fig. (\ref{fig2}a),
as well as in regions $O3$ and $O4$, classical thermal fluctuations dominate.
These regions can be construed as being quantum critical.
Further away in region $O1$,
the ferromagnetic phase is impervious to thermal fluctuations. The crossover to this regime
occurs when $\gamma\approx O(T)$.  This partition is the dashed
line separating region $O5$ from $O1$.  In Fig. (\ref{fig2}b), the
corresponding phases are shown for $r-r_c>0$.  The key difference
with the $r-r_c<0$ regime is the absence of the spin-glass phase.

Experimentally, the phase diagram has been measured for the random
Ising spin system $LiHo_xY_{1-x}F_4$ at finite temperature.  This
system possesses all three phases discussed here.  It then serves
as a bench mark test of the phenomenological theory we have
developed.  While the overall features of the
experimentally-determined phase diagram are similar to that shown
in Fig. (\ref{fig2}a), it is worth looking closely at the form of
the boundaries between the three phases.  Particularly striking in
the experimentally-determined phase diagram\cite{rosenbaum} is the
close to linear dependence of the PM/FM phase boundary away from
the bi-critical region but a non-linear dependence on the doping
level in the vicinity of the bi-critical region. This dependence
mirrors closely the behavior of the PM/FM finite temperature phase
boundary shown in Fig. (\ref{fig2}a).  While a quantitative
comparison cannot be made because of the phenomenological nature
of the coupling constants used in this model, the agreement with
experiment is sufficiently striking and serves to justify the
applicability of the model used here.

\subsection{$M>1$}

We consider now explicitly $M>1$.  For the problem at hand, the
ordered phase for $M=2$ corresponds to a superconductor. Analogous
isotropic solutions can be obtained for $M>1$ with the
transformations $u+v\rightarrow u+Mv$.  However, because non-zero mean
 generates spontaneously
an effective magnetization, there exists a possibility that the different
spin components of the replica Q-matrices might acquire
fundamentally different values as first proposed by Gabay and
Toulouse\cite{gt}. In the zero-mean case, this happens only when
a magnetic field is present. However, in this case, the
Gabay-Toulouse (GT) line exists for all $M>1$ as a result of the
spontaneously-generated magnetization.

To explore the possibility of a GT line, we must generalize the
ansatz for the Q-matrices to explicitly break the symmetry between
the spin-components of $Q$.  The simplest way of doing this is to
divide the spin components of the Q-matrix into longitudinal,
$\mu=\nu=1$ and transverse, $\mu=\nu\ne 1$ sectors. Hence, in Eq.
(\ref{qm}), we introduce the parameters, $q_L^{ab}$,
$\tilde{q}_L$, and $\bar {D}_L(\omega)$ for the longitudinal
$\mu=\nu=1$ component and $q_T^{ab}$, $\tilde{q}_T$, and $\bar
{D}_T(\omega)$ for the transverse components $\mu>1$.  At the
replica-symmetric level, both $q^{ab}_L$ and $q_T^{ab}$ are
constants independent of the matrix label, $ab$.  We will call
these constants, $q_L$ and $q_T$, respectively.  The resultant
expression for the free-energy \beq\label{fm1}
\frac{\cal{F}}{n}&=&\frac{1}{\beta\kappa t}\sum_{\omega\ne 0}
(\omega^2+r)\bar D_L(\omega)+\frac{M-1}{\beta\kappa
t}\sum_{\omega\ne 0}(\omega^2+r)\bar{D}_T(\omega)\nonumber\\ &&
+\frac{r}{\kappa t}\tilde{q}_L+(M-1)\frac{r\tilde{q}_T}{\kappa t}
 -\frac{\kappa}{3\beta
t}\sum_\omega\left(\bar{D}_L^3(\omega)+(M-1)\bar{D}_T^3(\omega)\right)+\nonumber\\
&&+\frac{u}{2t}\left\{\left[\tilde{q}_L+\frac{1}{\beta}\sum_{\omega\ne
0}\bar{D}_L(\omega)\right]^2+(M-1)\left[\tilde{q}_L+\frac{1}{\beta}\sum_{\omega\ne
0}\bar{D}_T(\omega)\right]^2\right\}\nonumber\\ &&
+\frac{v}{2t}\left\{\tilde{q}_L+(M-1)\tilde{q}_T+\frac{1}{\beta}\sum_{\omega\ne
0}\left(\bar{D}_L(\omega)+(M-1)\bar{D}_T(\omega)\right)\right\}^2\nonumber\\
&&-\frac{\kappa\beta^2}{3t}(\tilde{q}^3_L-3\tilde{q}_Lq_L^2+2q_L^3)-(M-1)\frac{\kappa\beta^2}{3t}
(q_T^3-3\tilde{q}_Tq_T^2+2q_T^3)\nonumber\\
&&-\frac{\beta\psi^2}{\kappa g
t}(\tilde{q}_L-q_L)+\frac{1}{2g}\left(-\gamma\psi^2+\frac{\zeta
\psi^4}{2}\right) \eeq
 is a generalization of Eq. (\ref{fed}) to
an anisotropic system.  The explicit factor of $M-1$ arises from
the separation into transverse and longitudinal components.

If we approach the GT line from below, we find that the
relevant saddle-point equations are
 \beq\label{mg1spe}
 \Delta_L^2&=&r+(u+v)\kappa[\tilde{q}_L+\frac{1}{\beta}\sum_{\omega\ne
 0}\bar{D}_L(\omega)]+(M-1)v\kappa[\tilde{q}_T+\frac{1}{\beta}\sum_{\omega\ne
 0}\bar{D}_T(\omega)]\nonumber\\
 \bar{D}_L(\omega)&=&-\frac{1}{\kappa}(\omega^2+\Delta_L^2)^{\frac12}\nonumber\\
 \bar{D}_T(\omega)&=&-\frac{1}{\kappa}|\omega|\\
 q_L&=&\frac{\psi^2}{2\kappa g\Delta_L}\nonumber\\
 \tilde{q}_L&=&\frac{\psi^2}{2\kappa g\Delta_L}-\frac{\Delta_L}
{\kappa\beta}\nonumber\\
 0&=&\psi\left(-\gamma+\zeta\psi^2+\Delta_L\right)\nonumber\\
 q_T&=&\tilde{q}_T=\frac{1}{\kappa}\left\{\frac{1}{\beta}\sum_\omega|\omega|+
\frac{1}{u+(M-1)v}\left(\frac{v}{\beta}
 \sum_\omega\sqrt{\omega^2+\Delta_L^2}-r-\frac{v\psi^2}{2g\Delta_L}
\right)\right\}\nonumber
 \eeq
which are a direct generalization of the $M=1$ equations to the
anisotropic system.  We are particularly interested in the
solution in the $\gamma$-$r$ plane where $q_T=0$.  This demarcates
the GT line. Below the GT line, $\tilde{q}_T=q_T\ne 0$, while
above, the transverse replica off-diagonal component of $Q$
vanishes. Note this state of affairs does not occur unless
$\psi\ne 0$.  If we substitute the non-trivial solution for $\psi$
into the expression for $\Delta_L^2$, we find that within
logarithmic accuracy at zero temperature, we recover the result
obtained previously for $M=1$ but with $u+v\rightarrow u+Mv$.  The
phase diagram hence is identical to that shown in Fig.
(\ref{fig1}). However, a new region, $\tilde{O}1$, appears.  This
is illustrated in Fig. (\ref{fig3}a). To find the line demarcating
this region we must solve for the transverse replica off-diagonal
component of $Q$. After several manipulations of the set of
equations in Eq. (\ref{mg1spe}), we find that \beq
q_T=\frac{1}{\kappa(u+Mv)}\left[r_c-r-\frac{v}{u}\Delta_L^2\right].
\eeq If we use the fact that $\Delta_L\approx\gamma$, we find that
the GT line occurs when \beq\label{gtline}
\gamma=\sqrt{\frac{u}{v}(r_c-r)}. \eeq The phase diagram depicting
this line at zero temperature is shown in Fig. (\ref{fig3}a). In
the region labeled $\tilde{O}1$, $q_T\ne 0$ and $q_L\ne 0$,
whereas in $O1$ only $q_L\ne 0$. Hence, we have identified the
zero-temperature GT line.  At finite temperature, the
generalization of the Eq.(\ref{gtline}) is simply \beq\label{ftgt}
\gamma=\sqrt{\frac{u}{v}\left[r_c(T)-r\right]}=\sqrt{\frac{v}{u}\left[r_c-r-(u+Mv)\frac{\pi
T^2}{3}\right]}. \eeq Hence, the GT line is now a surface in the
$\gamma$, $r$, and $T$ space, a slice of which is shown in Fig.
(\ref{fig3}b). At the point $\gamma=0$, we recover the isotropic
result that \beq q_T=q_L=\frac{r_c(T)-r}{\kappa(u+Mv)} \eeq in the
spin-glass phase.

We now consider the region above the GT line.  This region was not
analyzed by RSY.  However, this region is of considerable interest
because although $q_T$ vanishes in this region, the non-zero
transverse component of the order parameter becomes gapped,
significantly different in character from the longitudinal one.
This can be seen immediately by setting $q_T=\tilde{q}_T=0$ in Eq.
(\ref{fm1}) and differentiating with respect to
$\bar{D}_T(\omega)$.  From this operation, we find that contrary
to the ungapped transverse component of $Q$ below the GT line,
\beq\label{dt}
\bar{D}_T(\omega)=-\frac{1}{\kappa}(\omega^2+\Delta_T^2)^\frac12
\eeq with \beq
\Delta_T^2=r+\frac{u+(M-1)v}{\beta}\sum_\omega\sqrt{\omega^2+\Delta_L^2}+
v\left(\frac{\psi^2}{2g\Delta_L}-\frac{1}{\beta}
\sum_\omega\sqrt{\omega^2+\Delta_L^2}\right) \eeq above the GT
line.  The corresponding expression for $\Delta_L$ is easily
obtained from the first equation in Eq. (\ref{mg1spe}) by setting
$q_T=0$.  The ferromagnetic phase above the GT line can be divided
into two regions, $O1$ and $O2$, which are now different with
respect to the relative magnitudes of $\Delta_L$ and $\Delta_T$.
In the region $O2$ which is completely analogous to the
corresponding region in the $M=1$ case, $\Delta_L$ and $\Delta_T$
are almost equal and are given by Eq. (\ref{gamline}) with $u+v
\rightarrow u+Mv$. The condition for crossover to $O1$ in which
$\Delta_L$ and $\Delta_T$ are somewhat different in magnitude is
given by Eq. (\ref{gaml2}). This conclusion is reached by
manipulating the system of equations in Eq. (\ref{mg1spe}) with
$q_T=\tilde{q}_T=0$ and $\bar{D}_T(\omega)$ given by Eq.
(\ref{dt}). The resultant expression, \beq\label{eq53}
v\Delta_L^2-(u+v)\Delta_T^2=-ur+\frac{u(u+Mv)}{\beta}
\sum_\omega\sqrt{\omega^2+\Delta_T^2}, \eeq which is valid at any
temperature, contains both the transverse and longitudinal gaps.
Within the approximation that $\Delta_L\approx\gamma$, we obtain
that at zero temperature in $O1$, \beq\label{eq54}
\Delta_T=\left(\frac{4\pi(v\gamma^2+u(r-r_c))}{u(u+Mv)
\ln\frac{\Lambda_\omega^2}{(v\gamma^2+u(r-r_c))}}\right)^{\frac{1}{2}}
\eeq  From this equation it is immediately clear that $\Delta_T$
is logarithmically smaller than $\Delta_L$.  Also, we easily
recover the GT line, $\gamma=\sqrt{u(r_c-r)/v}$, simply by solving
$\Delta_T=0$.

At finite temperature, we can formally distinguish three limiting
cases: 1) $\Delta_L\gg T$ and $\Delta_T\gg T$, 2) $\Delta_L\gg T$
and $\Delta_T\ll T$, and  3) $\Delta_L\ll T$ and $\Delta_T\ll T$.
The regime $\Delta_L\ll T$ and $\Delta_T\gg T$ does not exist as
$\Delta_T$ is always less than $\Delta_L$. Case (1) is identical
to the $T=0$ limit whereas cases (2) and (3) are the
high-temperature limit with respect to $\Delta_T$.  To describe
(2) and (3), we calculate the sum in Eq. (\ref{eq53}) in the
high-temperature limit with the approximation that
$\Delta_L\approx\gamma$.  Within logarithmic accuracy, the
resultant equation \beq
v\gamma^2-u(r_c(T)-r)=u(u+Mv)\left(\frac{\Delta_T^2}{2\pi}
\ln\frac{\Lambda_\omega}{T}+T\Delta_T\right) \eeq is similar in
structure to Eq. (\ref{eq40}).  Consequently, within cases (2) and
(3), two distinct regimes denoted by $O5$ and $O5'$ and $O6$ and
$O6'$, respectively, arise. In the regions superscripted with a
prime, two conditions hold \beq \sqrt{v\gamma^2-(r_c(T)-r)}\ll
T,\quad\Delta_T=\frac{v\gamma^2-u(r_c(T)-r)}{T}\quad  O5\quad {\rm
and} \quad O6 \eeq Contrastly, in the unprimed regions, \beq
\sqrt{v\gamma^2-(r_c-r)}\ll T,\quad\Delta_T=\sqrt{\frac{2\pi
T^2}{3\ln\frac{\Lambda_\omega}{T}}}\quad  O5'\quad {\rm and} \quad
O6'. \eeq The transition between the primed and unprimed regions
occurs when $\sqrt{v\gamma^2-u(r_c(T)-r)}\approx
T/\ln\Lambda_\omega/T$, whereas the transition between $O6'$ and
$O1$ occurs when $\sqrt{v\gamma^2-u(r_c-r)}\approx T$. The
difference between regions $O5$ and $O5'$ and regions $O6$ and
$O6'$ is that $T\ll\Delta_L$ in the latter whereas the opposite is
true in the former.  We have taken particular care in
distinguishing the primed from the unprimed regions because they
imply that the GT transition is identical in structure to the
ordinary paramagnet/spin-glass transition described by RSY.
Except, only the transverse component of $Q$ is affected at the GT
transition. In fact, regions $O5'$ and $O6'$ are quantum critical
with respect to the GT transition while the temperature dependence
of the transverse component is `classical' in regions $O5$ and
$O6$. In $O1$, thermal fluctuations are subservient to quantum
fluctuations for both transverse and longitudinal components of
$Q$.

In each of these regions, a key question that can be addressed is
how does the transverse gap renormalize $\psi$ and $\Delta_L$.
Consider first the regime $\gamma\gg T$. In this regime, we find
that \beq\label{gtt2}
\Delta_L=\gamma-\zeta\psi^2(T)=\gamma-\frac{2g\zeta}{u+Mv}\left[{\gamma^3\over
2\pi}\ln\frac{\Lambda_\omega}{\gamma}
+(r_c-r)\gamma-\frac{(M-1)v\gamma}{u(u+Mv)}\Delta_T^2\right]. \eeq
It follows immediately from Eq. (\ref{eq54}) that the transverse
contribution to the longitudinal gap is logarithmically small. As
a result, the order parameter $\psi$ is well described by its
value given by Eq. (\ref{psi2}) with $u+v\rightarrow u+Mv$.
Consider now the high-temperature regime.  In this case,
$\Delta_L$ is exactly given by Eq. (\ref{ftpsi}) plus the term
proportional to $\Delta_T^2$ in Eq. (\ref{gtt2}).  However, we
checked explicitly that in every sub-regime, the correction due to
$\Delta_T^2$ is sub-dominant to the leading terms. Consequently,
the order parameter, $\psi$, and $\Delta_L$ are both
unrenormalized by the transverse gap to the leading order.

Using the solutions delineated in Eq. (\ref{mg1spe}), we can
calculate the free-energy density below and above the GT line.
Above and below the GT line, we find that the free-energy density
at zero-temperature up to the leading $\gamma-$ dependent terms,
\beq \frac{\cal{F}}{n}={\cal F}_0(u+v\rightarrow
u+Mv,\zeta)-\frac{(M-1)v\Lambda_\omega^2\gamma^2}{4\pi
u}+\frac{\gamma^2(r-r_c)}{2u(u+Mv)}\left(1-\frac{(M-1)v}{u}\right)+\cdots,
\eeq contains the standard analogous contribution from the $M=1$
analysis as well as $q_T$ and $\Delta_T$-dependent terms arising
from Eq. (\ref{fm1}). It is the contribution from the latter terms
that results in a suppression of the
$\gamma^4\ln\Lambda_\omega/\gamma$ as the leading
$\gamma-$dependent terms when $r=r_c$.  At the bi-critical point,
we find that in contrast to the M=1 case, the leading term in the
free energy density is analytic in the coupling constant,
$\gamma$.  This term is of the form, $\Lambda_\omega^2\gamma^2$.

\section{Replica-Symmetry Breaking}

The requisite\cite{gt} for a replica asymmetric solution within
Landau theory of spin glasses is the presence of a $Q^{4th}$ term
in the Landau action. We specialize to $M=1$ for simplicity.
To facilitate such an analysis, we must
extend the cumulant expansion of Eq. (\ref{free}) to the next
order in perturbation theory.  As RSY have performed such an
analysis for the spin-glass phase, we focus on the ferromagnetic case.
While several types of
fourth-order terms occur, the most relevant is of the form,
 \beq
-\frac{y_1}{6t}\int d^d x\int
d\tau_1d\tau_2\sum_{ab}\left[Q^{ab}(x,\tau_1,\tau_2)\right]^4
\eeq
This term will give rise to a $(q^{ab})^4$ contribution to the
free-energy.  Our focus is the resultant change in the free energy
\beq\label{deltaf}
\frac{\Delta\cal{F}}{n}=-\frac{\kappa\beta^2}{3t}\left(Tr
q^3+3\tilde{q}Tr q^2\right)-\frac{\beta\psi^2}{\kappa g
t}\sum_{a,b} q^{a,b}-\frac{y_1\beta}{6t}\sum_{ab}(q^{ab})^4.
\eeq

The presence of the $\psi^2$ term suggests that within the space
of ultrametric functions\cite{parisi} $q(x)$ on the interval $0\le
x\le 1$, we should choose an ansatz for $q$,
\beq\label{qs} q(s)= \left\{ \begin{array}{ll}
     q_0 & \quad 0<s<s_0 \\
     \frac{\kappa\beta s}{2y_1} & \quad s_0<s<s_1 \\
     q_1 & \quad s_1<s<1
\end{array}\right.
\eeq
 that has two distinct plateaus.  This insight is based on an
analogy with the replica broken-symmetry solution in the presence
of a magnetic field.  From continuity, we must have that
$q_0=\kappa\beta s_0/2y_1$ and $q_1=\kappa\beta s_1/2y_1$.  The
constants $s_0$ and $s_1$ can be determined from the saddle-point
equations for $\Delta\cal{F}$.  Upon differentiating with respect
to $q_1$, we find that $\tilde{q}=q_1-y_1q_1^2/\kappa\beta$.  The
corresponding equation for $q_0$ provides a relationship \beq
q_0=\left(\frac{3\psi^2}{4y_1\kappa g}\right)^\frac13 \eeq between
$q_0$ and $\psi$.  Replica symmetry breaking occurs when $q_0<
q_1$. To leading order in temperature, we can approximate
$q_1\approx \tilde{q}$, where $\tilde{q}$ is given by \beq
\tilde{q}=\frac{\psi^2}{2\kappa g\Delta}. \eeq Hence, the boundary
demarcating the replica-symmetric solution is determined by \beq
\psi^\frac43=\left(\frac{6\kappa^2 g^2}{y_1}\right)^\frac13\Delta
\eeq which we obtain upon equating $q_0$ and $q_1$.  If we use Eq.
(\ref{psi2}) for $\psi$ which is valid in region $O1$ and use the
fact that in this region $\Delta\approx \gamma$, we obtain \beq
\gamma=\frac{2 y_1(r_c-r)^2}{3\kappa^2 g(u+v)^2} \quad r<r_c \eeq
as the condition for replica symmetry breaking at low
temperatures, $T \ll\gamma$. This condition for replica-symmetry
breaking extends continuously to the spin-glass phase agreeing
with the work of RSY.  The phase diagram illustrating the
replica-broken symmetry region is depicted in Fig. (\ref{fig5}).
 At finite temperature, we use
Eq. (\ref{ftpsi}) for $\psi$ and obtain a generalization,
 \beq\label{rsbft}
  \gamma=\frac{2 y_1}{3\kappa^2 g}\left[\gamma
T+ \frac{r_c-r}{u+v}\right]^2 \quad r<r_c, \eeq of the
replica-breaking condition which extends smoothly over to the
zero-temperature condition.

To estimate the strength of the replica-symmetry breaking, we
define the effective broken ergodicity order parameter, \beq
\Delta_q=q_1-\int_0^1 q(s)ds. \eeq If we substitute the expression
for $q(s)$ and integrate, we obtain \beq
\Delta_q=\frac{y_1T}{\kappa}(q_1^2(T)-q_0^2(T)).
 \eeq
Because
both $q_0$ and $q_1$ are finite at low temperatures,
 $\Delta_q\rightarrow 0$ as
$T\rightarrow 0$.  The weakness of the replica-symmetry breaking
in the ferromagnetic phase is in accord with the weak-symmetry
breaking found by RSY in the spin-glass phase.  Implicit in the
replica-broken solution in Eq. (\ref{qs}) is the presence of many
degenerate energy minima in the energy landscape.  The weakness of
replica symmetry-breaking at low temperatures in this model within
the ferromagnetic phase suggests that the ferromagnetic phase is
energetically homogeneous.

\section{Summary}

We have constructed here a Landau theory near the bi-critical
point for a ferromagnet, spin-glass, and paramagnet.  The
analogous analysis was also performed for $M>1$ in which the
ordered phase for $M=2$ is a superconductor. All transitions were
found to be second order in contrast to the assertion by Hartman
and Weichman\cite{nzm} who claimed that the FM/PM transition was
first order in the spherical limit.  A key point of our analysis
is the formal equivalence between the role of a non-zero mean and
the presence of a magnetic field in the zero mean problem.  This
observation is equally valid for classical systems. Resilience of
the ferromagnet against thermal fluctuations occurs when
$T<\gamma$, where $\gamma$ is the coupling constant that
ultimately determines the rigidity of the ferromagnet phase.
Additional features of our analysis that are particularly striking
are 1) the non-analytic dependence of $\psi$ on $\gamma$, namely
$\psi\propto\gamma^{3/2}/\ln\Lambda_\omega/\gamma$,
 near the bi-critical region, in the
vicinity of the bi-critical point and 2) the subsequent
leading non-analytic dependence of the free-energy density in region $O1$
on $\gamma$ for $M=1$.  We showed that this behavior ceases for $M>1$ below
and above the GT line.  The reason underlying this difference with
the $M=1$ case is the presence of $M-1$ components of $Q$ that yield
leading analytic contributions of order $O(\gamma^2)$ to the free energy
density.  An additional feature which our analysis brings out for $M>1$ is
the similarity between the Gabay-Toulouse transition with the spin-glass/paramagnet
transition with zero mean. This similarity has been noticed previously
in the context of classical spin-glasses\cite{by}.  A key surprise found
in the analysis of the GT transition is the
sub-leading depdence of $\psi$ and $\Delta_L$ on the transverse
gap and $q_T$.  This suggests that the GT line
should have only weak experimentally-detectable features in the
superconducting phase, for example.  The excellent agreement observed with the
experimental results on $LiHo_xY_{1-x}F_4$ for the case of $M=1$ is encouraging
that similar agreement will be found with experiments on
the analogous superconducting systems.

While we referred to the point of intersection between all the
phases as being bi-critical, it is in fact multi-critical. This
state of affairs obtains as a result of the presence of replica
symmetry breaking and the Gabay-Toulouse instability. As in the
$M=1$ spin-glass case, we also showed that the non-ergodicity
parameter is linear in temperature illustrating the weakness of
replica symmetry breaking in the ferromagnetic phase.  Though we
did not treat explicitly replica-symmetry breaking for $M>1$, such
symmetry breaking is expected in this case as well when quartic
terms are included in the action.  In a future study, we will
extend this analysis beyond mean field and report on the
renormalization group analysis of the `bi-critical' region.

\acknowledgements We thank S. Sachdev for useful comments during
the early stages of this project.  This work was funded by the DMR
of the NSF and the ACS petroleum research fund.

\begin{figure}
\caption{M=1 finite-temperature phase diagrams for a) $r_c-r>0$
and b) $r-r_c>0$. Regions $O3$ and $O4$ lie close to the
paramagnet (PM) boundary and hence have $\gamma\approx \Delta(T)$.
Thermal fluctuations dominate in region $O5$ as well as in $O3$
and $O4$.  In region $O1$, thermal fluctuations are negligible.
The key difference between noted when $r_c-r$ changes sign from
positive to negative is the absence of the spin-glass (SG) phase
for $r>r_c$.}
 \label{fig2}
 \end{figure}
 \begin{figure}
\caption{a) Zero and b) finite-temperature phase diagrams for
$M>1$.  The criteria for distinguishing regions $O1$ and $O2$ are
identical to the $M=1$ case but except $u+v\rightarrow u+Mv$. The
GT line separates regions $\tilde{O}1$ from $O1$. This line
terminates at $\gamma_1=\sqrt{u(r_c-r)/v}$.
 Below this line both the transverse and longitudinal
components of the replica off-diagonal components of the Q-matrix
are non-zero.  The location of the GT line is given by Eq.
(\protect\ref{gtline}).  b) Finite temperature GT line as
determined by Eq. (\protect\ref{ftgt}).} \label{fig3}
\end{figure}
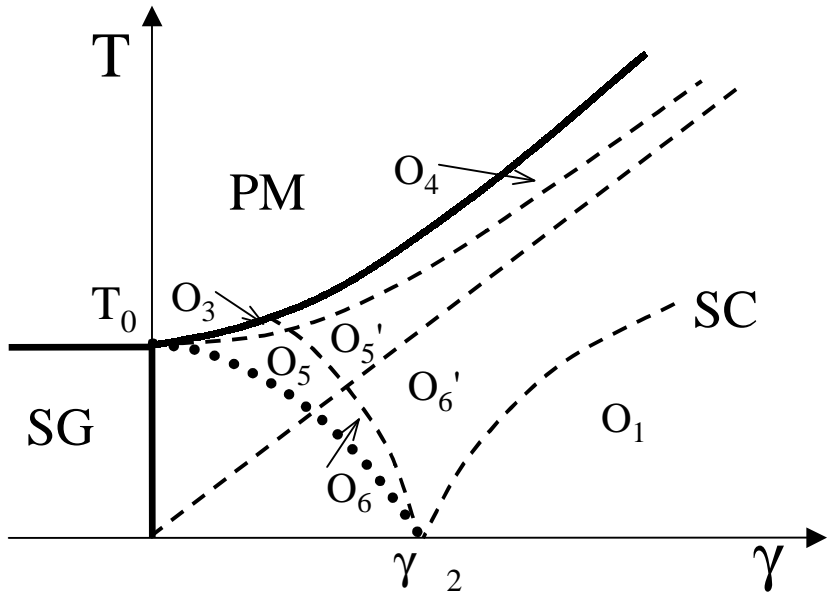
\begin{figure}
\caption{Phase diagram illustrating the distinct regimes that
occur at finite temperature above the GT line denoted with solid
circles. Regions $O3$ and $O4$ are as described in Fig.
(\protect\ref{fig2}a) previously.  Regions $O5'$ and $O6'$ quantum
critical with respect to the GT transition. The difference between
regions $O5$ and $O5'$ and regions $O6$ and $O6'$ is that
$T\ll\Delta_L$ in the latter whereas the opposite is true in the
former.} \label{fig4}
\end{figure}
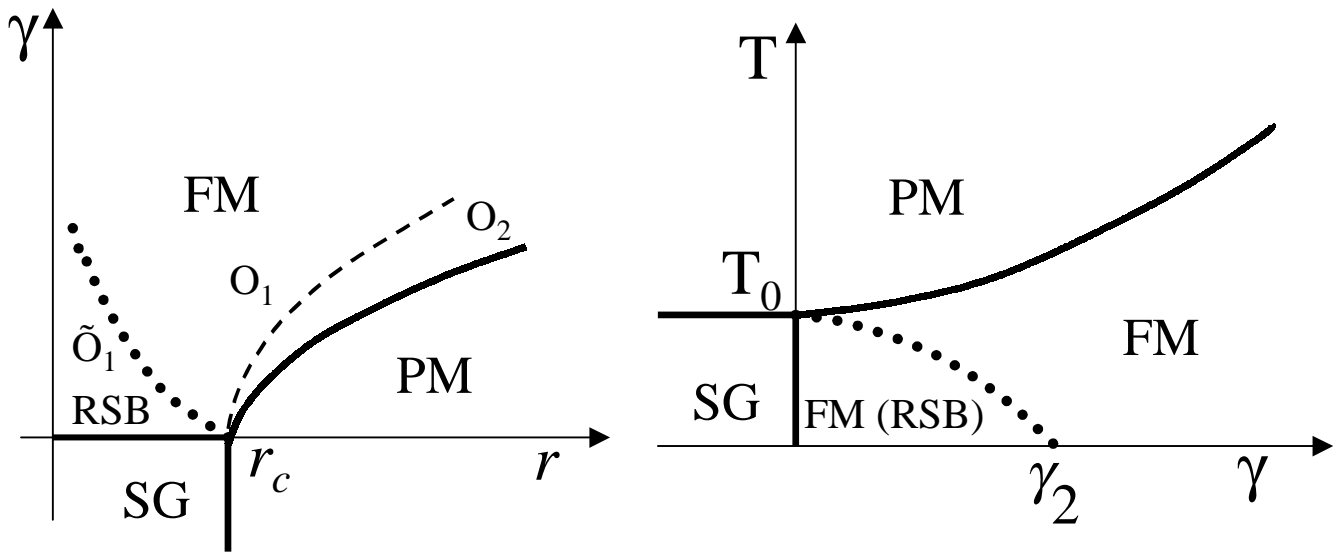
\begin{figure}
\caption{a) Phase diagram illustrating replica-symmetry breaking
for M=1 in the $\gamma-r$ plane.  The broken symmetry region is
denoted RSB. Regions $O1$ and $O2$ are as before. b)
Replica-symmetry breaking region (shaded region) at finite
temperature. The finite temperature criterion is given by Eq.
(\protect\ref{rsbft}).  The point
$\gamma_2=2y_1(r_c-r)^2/(3\kappa^2 g(u+v))$ is determined by the
replica-symmetry breaking condition specified in Eq.
(\protect\ref{rsbft}).}
 \label{fig5}
\end{figure}
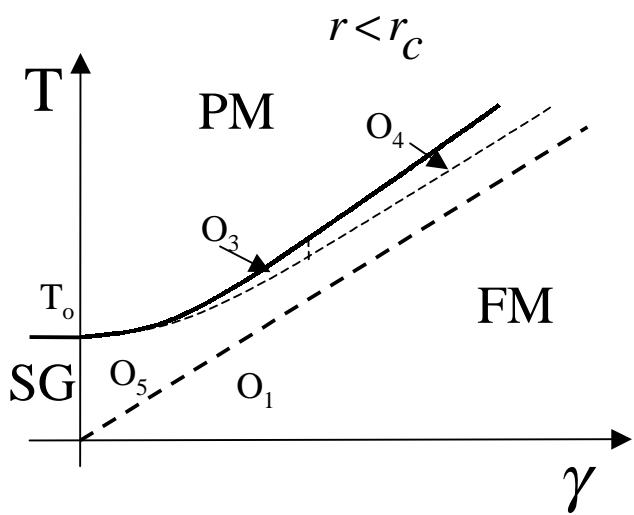
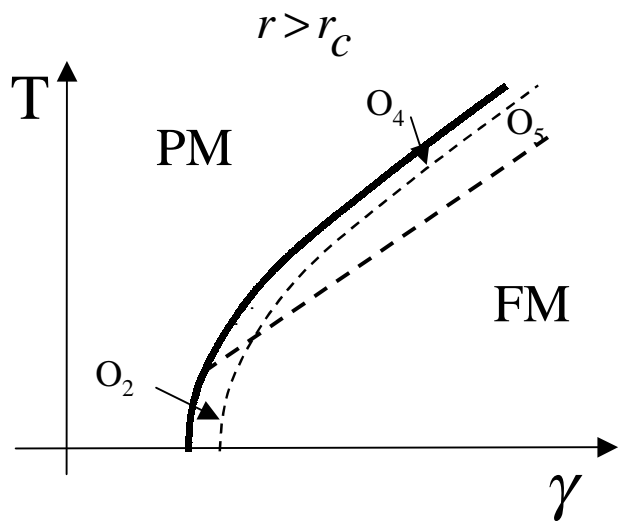
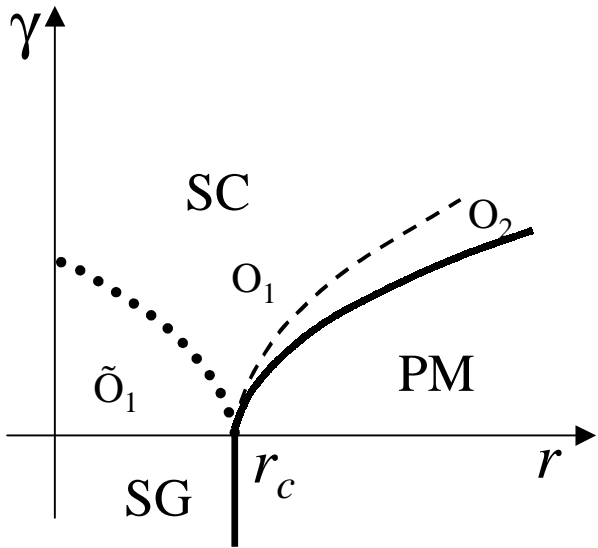
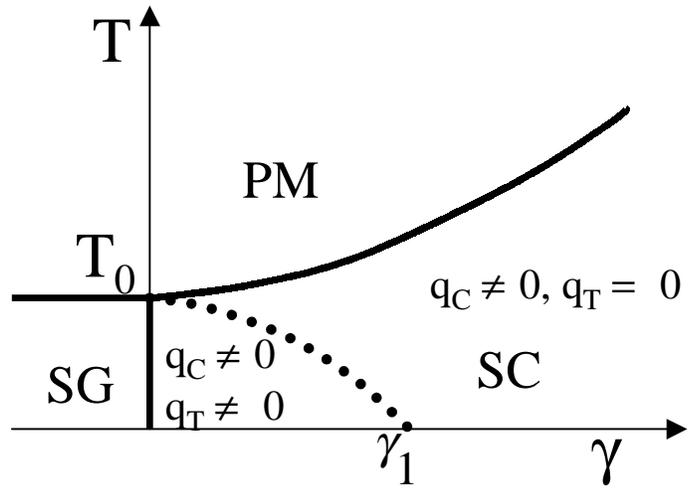
\end{document}